\documentclass[sigconf, 10pt]{acmart}
\usepackage{graphicx} % Required for inserting images
\usepackage{algorithmic}
\usepackage{graphicx}
\usepackage{url}
\usepackage{booktabs}
\usepackage{nameref}  % In preamble
\usepackage{float}  % Add this to your preamble
\usepackage{xcolor}

\title{MemorySim: An RTL-level, timing accurate simulator model for the Chisel ecosystem }
\author{Ansh Chaurasia}
\date{May 2025}

\begin{document}

\maketitle

\section{Abstract}
AI applications have seen a surge in demand, leading to an uptick in attention towards AI chips. These applications have created a myriad of opportunities in hardware space. One area of opportunity for all hardware systems is the memory subsystem, which often, through bandwidth and performance, dictates system performance of high-demand applications like large language models (LLMs). 

Currently there is a dire limitation in register-transfer level (RTL) memory subsystem models with support for obtaining profiling data. High-level memory models like DRAMSim2 \cite{yu2021dramsim3} and DRAMSim3 \cite{weber2010dramsim2} exist for simulations, but implementations like these often do not support both \textit{timing} and \textit{correctness}. 

We introduce MemorySim, a RTL-level memory simulator that strives to provide accurate timing simulations of memory systems while retaining correctness. The simulator is designed to be integrated into existing RTL-level simulations written in Chisel or Verilog. It's also compatible with the Chisel \cite{bachrach2012chisel} / Chipyard \cite{waterman2020chipyard} ecosystem, allowing users to obtain highly accurate representations of performance and power through direct integration downstream leverage of simulation tools such as FireSim \cite{zhang2020firesim}.   

\section{Introduction}

The advent of the Transformer architecture has revolutionized sequence modeling by replacing recurrent and convolutional networks with self-attention mechanisms, enabling highly parallelizable models that achieve state-of-the-art results across a wide range of tasks \cite{vaswani2017attention}. Building upon this foundation, the paradigm of \emph{foundation models}—large pre-trained models adaptable to diverse downstream tasks—has emerged, exemplified by GPT-3’s 175 billion-parameter few-shot capabilities \cite{brown2020language} and the broad taxonomy of foundational systems spanning language, vision, and multimodal reasoning \cite{bommasani2021opportunities}. Scalability studies further characterize the emergent capabilities and performance trends of these models under increasing compute and data budgets \cite{kaplan2020scaling}.

Despite algorithmic innovations, the memory subsystem of modern accelerators often becomes the primary bottleneck in both training and inference of large language models (LLMs). Techniques such as the ZeRO optimizer stages mitigate memory redundancies and extend feasible model sizes by partitioning optimizer state, gradients, and parameters across devices \cite{rajbhandari2020zero}. Yet recent GPU-level analyses reveal that inference—especially under large batch sizes—remains predominantly memory-bound due to DRAM bandwidth saturation, hindering compute utilization even on high-end hardware \cite{recasens2025mind}.  

During training, large-batch regimes and activation check-pointing for model parallelism elevate peak memory demands, driving complex combinations of data, tensor, and pipeline parallelism to avoid out-of-memory failures \cite{kundu2024performance}. For inference, the quadratic complexity of self-attention and auto-regressive generation exacerbate cache and bandwidth pressures, prompting system-level optimizations such as kernel fusion and heterogeneous offloading \cite{ning2024survey}. These observations underscore the need for RTL-level memory simulation and profiling tools that can deliver accurate timing and correctness information to guide hardware-aware architecture exploration.

\section{Current Limitations}

While high‐level DRAM models such as DRAMSim2 \cite{weber2010dramsim2} and DRAMSim3 \cite{yu2021dramsim3} have become de facto tools for architectural exploration, they face a core triad of limitations when applied to AI‐accelerator memory subsystem analysis:  

\begin{enumerate}
    \item \emph{Timing fidelity}: Both DRAMSim2 and DRAMSim3 use simplified cycle abstractions that approximate row‐buffer hits, precharges, and refresh behavior, but omit critical microarchitectural effects—such as bank‐slice contention and internal command reordering—which can skew effective bandwidth and latency estimates under heavy, burst‐oriented traffic common to transformer inference \cite{kim2015ramulator}. 

    \item \emph{Resource constraints}: Without RTL‐accurate description of resources, simulators struggle to capture limitations beyond a behavioral abstraction. Abstractions such as \textit{backpressure}, and cycle based limitations
    
    cannot guarantee bit‐true data integrity or capture corner‐case hazards (e.g., tRRD violations under concurrent multi‐bank accesses), making them unsuitable for early integration tests of new memory‐centric accelerator designs \cite{arnold2015validating}. 

    \item \emph{Power–performance coupling}: Standalone power‐estimation frameworks such as DRAMPower \cite{schellekens2014dram} and VAMPIRE \cite{ardakani2017vampire} often rely on cycle‐stack traces produced by DRAMSim2 or gem5; yet the loose coupling between timing and energy models precludes accurate estimation of dynamic power under throttling or command prioritization schemes optimized for AI workloads \cite{shevgoor2016understanding}.
\end{enumerate}

More cycle‐accurate DRAM frameworks—Ramulator \cite{kim2015ramulator}, USIMM \cite{mosbeck2015usimm}, and DRAMSys \cite{kai2018dramsys}—offer enhanced configurability and support for emerging DDR standards, but remain difficult to embed within register‐transfer‐level (RTL) simulations. Their C++ incarnations introduce foreign‐language interfacing overheads when integrated with Chisel or SystemVerilog based designs, resulting in slow co‐simulation or the need for custom wrappers that further erode timing validity \cite{harrison2020midas}.

Consequently, there exists a pressing need for an \emph{RTL‐native} memory simulator that: (a) exposes fine‐grained timing callbacks and decoupled command channels, (b) preserves data correctness via cycle‐accurate handshake protocols, and (c) seamlessly interoperates with hardware generators and FPGA‐accelerated emulation flows. By operating entirely within the Chisel \cite{bachrach2012chisel} / Chipyard \cite{waterman2020chipyard} ecosystem—leveraging the same FIRRTL intermediate representation and MIDAS‐based back‐ends \cite{harrison2020midas, zhang2020firesim}—such a simulator can accurately profile bandwidth, latency, and power consumption without sacrificing integration ease or scale.

\section{Project Targets}
Implement an RTL-level memory model in Chisel that can be accessed via Chipyard and FireSim emulations to estimate timing and power. As mentioned later in this paper, we plan to evaluate the fidelity of the model through rough comparisons with ideal memory systems.

\section{Simulator Architecture}

\subsection{Top-Level Architecture}

\begin{figure}[htbp]
    \centering
    \includegraphics[width=1\linewidth]{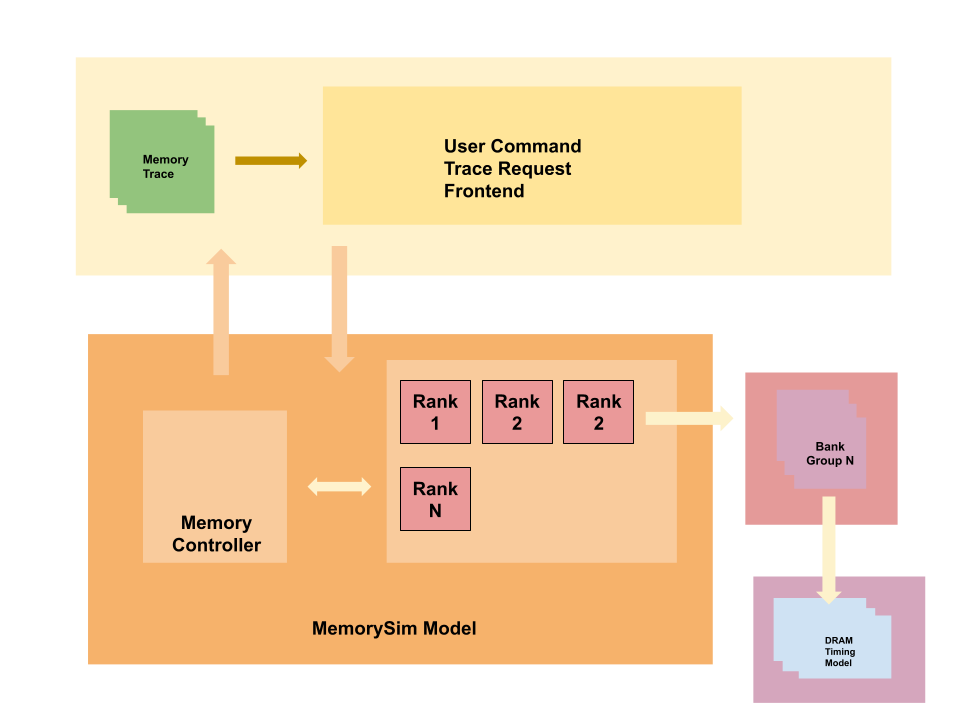}
    \caption{High-level architecture overview of the MemorySim simulation architecture. }
    \label{fig:top-level-diagram}
\end{figure}
\autoref{fig:top-level-diagram} showcases a high level overview of the various system components, and potential avenues for the user to test the memory model in a standalone fashion. MemorySim was designed to be interoperable with existing Chisel hardware, while also providing useful value through standalone, trace-based simulations. We provide verified options for the module to be integrated into existing Chisel repositories, but focus on the standalone executable for isolated experimentation and understanding. 

The core MemorySim architecture consists of the command front-end to receive standalone memory traces, which is received by a backend consisting of a memory controller and a memory channel. A typical path of a memory request would look like the following:
    
\begin{enumerate}
    \item Memory trace lists request \(R = \{\mathbf{addr},\,t\}\), where \(t\) is the cycle at which the request must be issued.
    \item At cycle \(t\), the request is enqueued into the global \texttt{reqQueue}. If the controller is not back-pressured, the request is dispatched in the very next cycle.
    \item The \texttt{MemoryController} classifies the request by rank and bank, forwards it into the appropriate Bank Scheduler’s local queue, and begins the ACTIVATE – READ/WRITE – PRECHARGE handshake.
    \item Each Bank Scheduler enforces closed-page policy and refresh deadlines by driving the downstream DRAM Timing Model, which holds the request in timing-parameter states (e.g.\ \(t_{\text{RCD}}\), \(t_{\text{RP}}\), \(t_{\text{RFC}}\)).
    \item Upon completion, the Bank Scheduler emits a \(\langle\text{data/ack}\rangle\) token back to the controller, which collects it in \texttt{respQueue} and returns a final acknowledgment to the frontend.
\end{enumerate}

\subsection{Bank Scheduler Design Architecture}
\begin{figure}[htbp]
    \centering
    \includegraphics[width=1\linewidth]{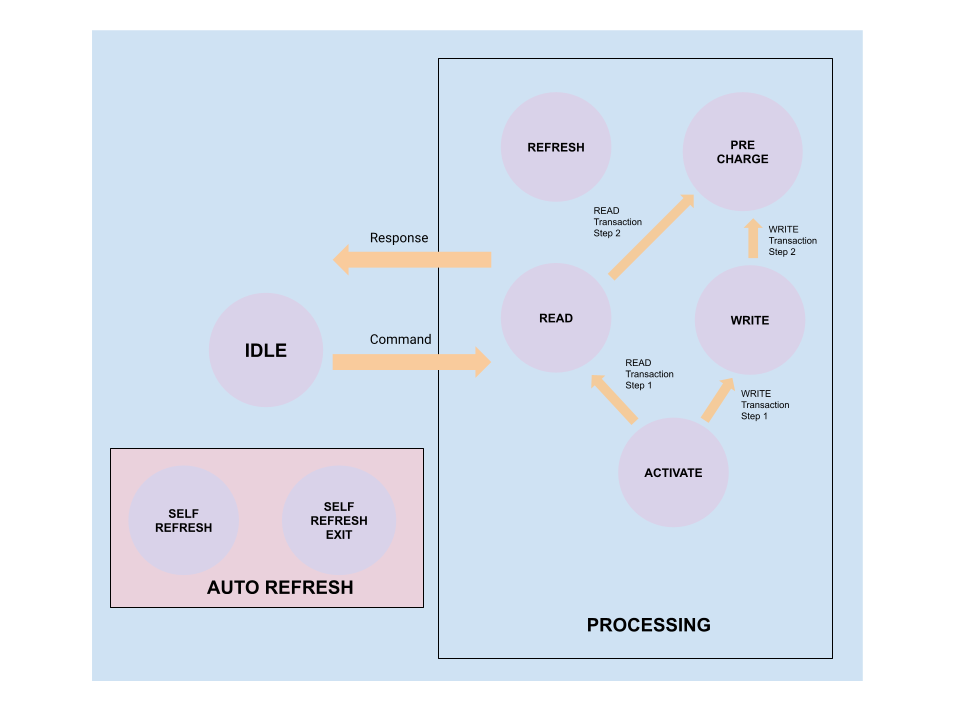}
    \caption{A diagram of finite-state-machine (FSM) orchestration of the bank scheduler. }
    \label{fig:bank-scheduler}
\end{figure}
A critical unit of the memory subsystem is the bank scheduler. This module interacts with the physical memory module. Each bank scheduler is designed to operate and manage a bank within the memory unit. Addresses are directly mapped to banks via a fixed address mapping. The address mapping scheme is shown below:

\[
\mathbf{address} \longleftarrow \{\mathbf{remaining bits}, \mathbf{rank idx}, \mathbf{bankgroup idx}, \mathbf{bank idx}\}
\]

By extracting the the $\mathbf{rank}$, $\mathbf{bankgroup}$, and $\mathbf{bank}$ bits, we can map an address to a specific DRAM bank scheduler.

Our DRAM Bank follows a \textit{closed-page} policy, which means every request, irrespective of its successors, are preceded by \texttt{ACTIVATE} and succeeded by \texttt{PRECHARGE}. As will be noted in our results in \autoref{sec:results-overview}, we discover that the ideal reference simulator \textit{always} uses an open page policy (despite attempts to change the configuration to use a closed policy), contributing to the total steady state cycle-difference.

\autoref{fig:bank-scheduler} shows a model of the various states employed by the closed-page bank finite-state machine at a high level. The states are divided into \texttt{IDLE}, \texttt{PROCESSING}, and self refresh related states used to park the DRAM bank during \texttt{IDLE} activity. The life-cycle transition paths for the scheduler are listed below.

\subsubsection{READ REQUEST}:
\begin{enumerate}
    \item Request initially is in \texttt{IDLE}
    \item If not within the refresh deadline, the scheduler issues an \texttt{ACTIVATE} request, powering on the DRAM row
    \item After an \texttt{ACTIVATE} completion acknowledgment (\texttt{ACTIVATE}-ack) is returned from the DRAM, the scheduler issues a READ request.
    \item After the \texttt{READ}-ack is returned, the scheduler issues a \texttt{PRECHARGE} command
    \item  Post \texttt{PRECHARGE}-ack, scheduler returns to IDLE, awaiting the next request.
\end{enumerate}

\subsubsection{WRITE REQUEST}:
\begin{enumerate}
    \item Request initially is in IDLE. 
    \item If not within the refresh deadline, the scheduler issues an ACTIVATE request, powering on the DRAM row
    \item After an \texttt{ACTIVATE} completion acknowledgment (\texttt{ACTIVATE}-ack) is returned from the DRAM, the scheduler issues a WRITE request.
    \item After the \texttt{WRITE}-ack is returned, the scheduler issues a \texttt{PRECHARGE} command
    \item  Post \texttt{PRECHARGE}-ack, scheduler returns to IDLE, awaiting the next request.
\end{enumerate}

\subsubsection{REFRESH REQUEST}:
\begin{enumerate}
    \item Request initially is in \texttt{IDLE}
    \item If within the refresh deadline, a REFRESH request is issued. 
    \item After a \texttt{REFRESH}-ack is returned from the DRAM, the scheduler returns to IDLE.
\end{enumerate}

The refresh request flow also supports self refreshes - if the bank has been idle for $1,000$ cycles, a self refresh request is issued, and upon an acknowledgement, moves into the \texttt{SREF ENTER} state. Upon a valid request arriving, it remains in that state. It then issues an \texttt{SREF EXIT} command to exit self refresh mode, and processes the next request.

\subsection{Memory Controller Architecture Design}

\begin{figure}[htbp]
    \centering
    \includegraphics[width=1\linewidth]{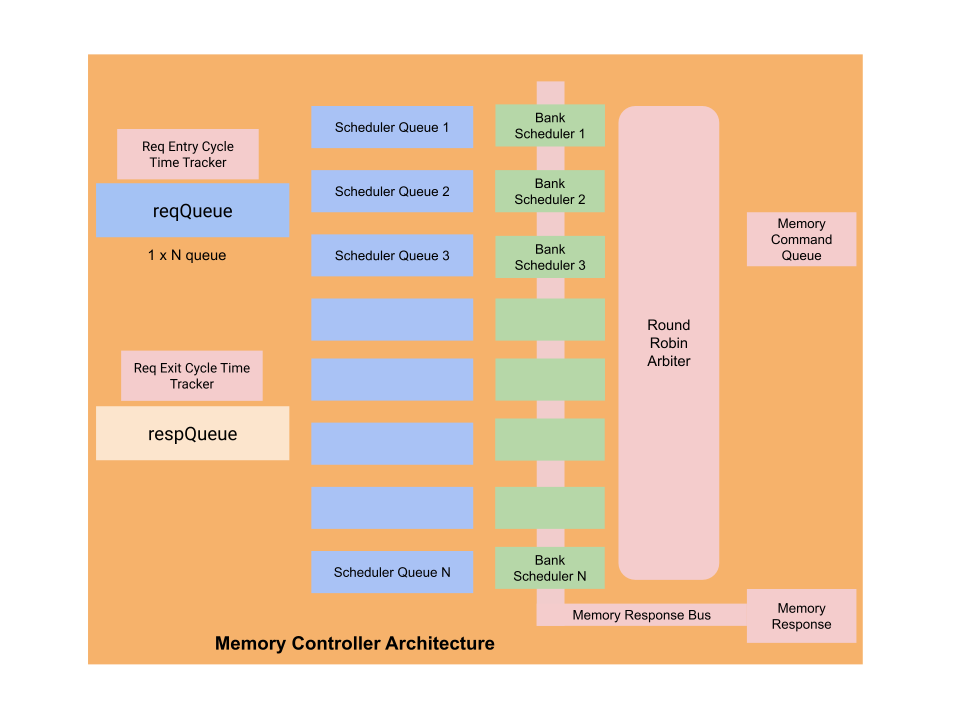}
    \caption{Diagram of the memory controller at a high level. The reqQueue data structure comprises of all the blue strips, and splits incoming requests into the respective scheduler queues. Once the scheduler receives responses from physical memory, the data is carried forth towards the response queue.}
    \label{fig:mem-controller-architecture}
\end{figure}
The memory controller is designed to issues requests with parallelism at the bank level. As shown in \autoref{fig:mem-controller-architecture}, requests enter through a request queue \textit{reqQueue}, and are issued back to the user through a response queue \textit{respQueue}. 

Intermediate memory commands from bank schedulers are sent from the bank scheduler to a round robin arbiter, which collects the requests, and enqueues them into the memory command queue. Responses from the physical memory channel are sent back to the controller, which broadcasts the response to all bank schedulers. The bank scheduler who's address mapping and request-id meet the response's will accept the request.

\subsection{Physical Memory Model Architecture}
\begin{figure}[htbp]
    \centering
    \includegraphics[width=1\linewidth]{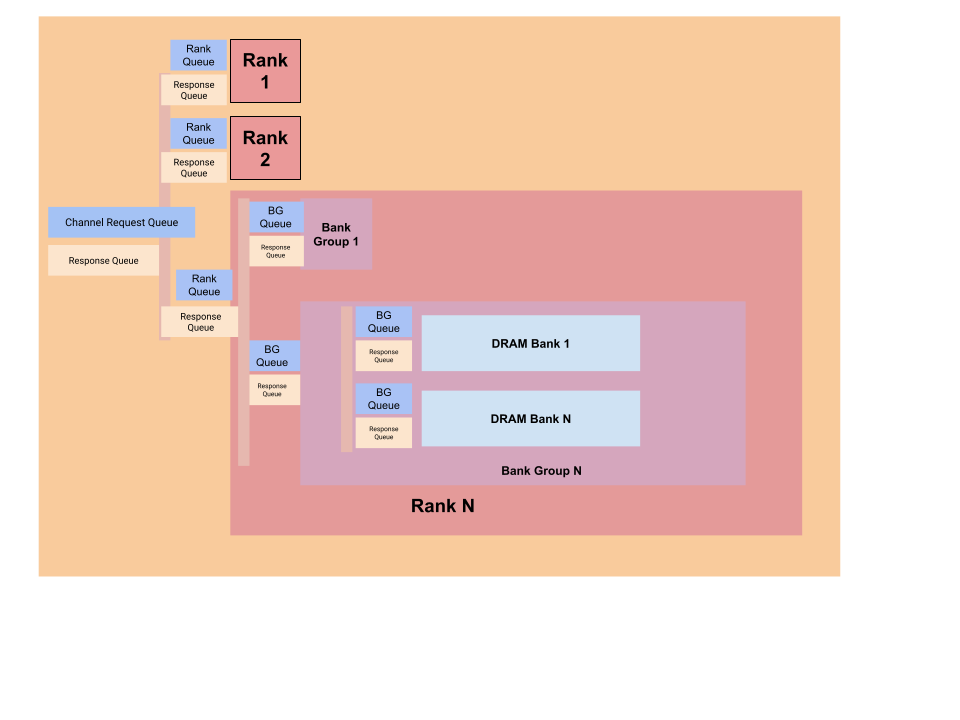}
    \caption{Top-level diagram of the memory hierarchy, which goes from (1) channels, (2) to ranks, (3) to bank groups, (4) to banks. The red bars denote RR arbiters, with the inputs solely being the light response queues. The responses are coalesced into the the broader response queue. No request queues interact with the round-robin arbitration.}
    \label{fig:physical-memory-top-level-architecture}
\end{figure}
The controller interacts with a memory channel, a diagram of which is shown above in \autoref{fig:physical-memory-top-level-architecture}. Each level of the hierarchy is separated by request and round-robin arbitrated response queues. The request queues leverage a similar multi-dequeue data structure to that discussed in the memory controller design. Functionally, the channel, rank, and bank group are indistinguishable, representing layers of the memory hierarchy. Our DRAM timing model breaks this tradition, representing the fundamental unit of memory storage, and as such requires more complex logic.

\subsection{DRAM Timing Model}

\begin{figure}[htbp]
    \centering
    \includegraphics[width=1\linewidth]{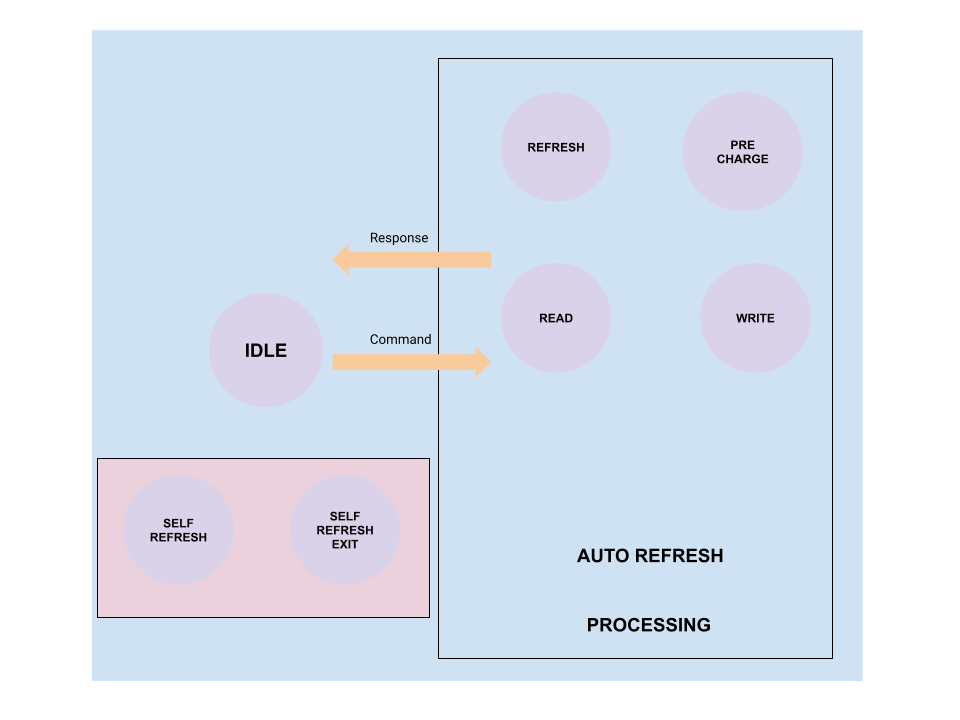}
    \caption{Diagram of the timing model used for the DRAM memory bank. The top-level diagram, although similar to that of the bank scheduler is different in function - no processing states are explicitly linked (i.e READ comes after PRECHARGE), rather these states are entered through control from the bank states.}
    \label{fig:dram-bank-model}
\end{figure}

The DRAM model bears many similarities, and can thought of as a mirror to the implementation of bank-scheduler FSM. This is by design. However, it is designed to be controlled by the bank scheduler and track row information, rather than make significant independent logic. Given a set of timing parameters, the model enters a state of \texttt{READ}, \texttt{WRITE}, \texttt{REFRESH}, \texttt{PRECHARGE}, \texttt{SELF REFRESH}, \texttt{SELF REFRESH EXIT}, waits for duration specified by a timing parameter, and then issues a response.

A list of timing parameters that are supported have been listed here - \autoref{tab:timing-parameter-values}
.

\begin{table}[htbp]
\centering
\begin{tabular}{c c p{2.5cm} c}
\toprule
\textbf{Parameter Name} & \textbf{Stage} & \textbf{Purpose} & \textbf{Value} \\
\midrule
tRP     & PRECHARGE & Total time it takes for a row pre-charge before activation & 14 \\
tFAW    & ACTIVATE  & Four activation window length – capped at 4 activations within this duration & 30 \\
tRRDL   & ACTIVATE  & Minimum cycles between two consecutive activates & 6 \\
tRCDRD  & READ      & Delay from activate to read & 14 \\
tCCDL   & READ      & Gap between consecutive reads & 2 \\
tWTR    & READ      & Turnaround time between write and subsequent read & 8 \\
tRP     & READ      & Time since last precharge & 14 \\
tRCDWR  & WRITE     & Delay from activate to write & 14 \\
tCCDL   & WRITE     & Gap between consecutive writes (depends on previous op) & 2 \\
tRP     & WRITE     & Time from precharge to write & 14 \\
tRFC    & REFRESH   & Deadline to start a refresh & 260 \\
tREFI   & REFRESH   & Interval between periodic refreshes & 3600 \\
\bottomrule
\end{tabular}
\caption{Timing parameters implemented by the bank model.}
\label{tab:timing-parameter-values}
\end{table}

\section{Experimentation Methodology}

\subsection{Introduction to DRAMSim3}

DRAMSim3 is a cycle-accurate, thermal-capable DRAM simulator supporting a wide range of DRAM protocols—including DDR3, DDR4, LPDDR3, LPDDR4, GDDR5, GDDR6, HBM, HMC, and emerging non-volatile memories—designed for both trace-driven and full-system simulation environments.   
It is implemented in modern C++ as an object-oriented framework, featuring parameterized DRAM bank models, modular memory controllers, command queues, and flexible system-level interfaces for integration with CPU simulators such as gem5 and ZSim or for standalone trace workloads.  
The core simulation kernel is a discrete event-driven scheduler that processes DRAM commands (\texttt{ACTIVATE}, \texttt{READ}, \texttt{WRITE}, \texttt{PRECHARGE}, \texttt{REFRESH}) according to timing parameters derived from JEDEC standards, ensuring cycle-accurate fidelity.  
DRAMSim3 optionally incorporates thermal and power modeling by coupling its event traces with energy-estimation engines, enabling detailed power–performance trade-off studies under throttling or scheduling policies optimized for AI workloads.  
The simulator is engineered for portability and parallelism: it compiles on Linux, macOS, and Windows, and it can leverage multiple threads to accelerate large trace-driven analyses while preserving accuracy.  

\subsection{DRAMSim3 Key Features}

\begin{itemize}
  \item \textbf{Modular Hierarchy}: Channels, ranks, bank groups, and banks are each represented by dedicated C++ classes, with timing parameters (e.g., \texttt{tRCD}, \texttt{tRP}, \texttt{tRAS}) loaded from JSON configuration files to match specific DRAM standards. 
  \item \textbf{Command Scheduling}: The built-in memory controller supports configurable scheduling policies (e.g., FCFS, FRFCFS), with per-bank command queues and arbitration logic to model realistic contention and bank-level parallelism. 
  \item \textbf{Trace and Full-System Modes}: Users can drive DRAMSim3 with simple address-trace files (request cycle, address, read/write opcode) or via hot-plug interfaces to full-system simulators, enabling evaluation from microbenchmarks to complex OS workloads.   
  \item \textbf{Refresh and Self-Refresh}: Periodic and on-idle refresh operations are modeled according to user-specified \texttt{tREFI} and \texttt{tRFC} parameters, including power-down self-refresh modes for low-power analysis.   
  \item \textbf{Parallel Execution}: A thread-pool approach partitions large trace files across worker threads, reducing wall-clock simulation time while maintaining correct event ordering.   
\end{itemize}

\section{Target Benchmarks}

These microbenchmarks were created to generate real-world memory access traces (via Valgrind) for evaluating MemorySim’s timing accuracy and correctness under representative workloads.

\begin{itemize}
  \item \texttt{conv2d.c}: A small 2D convolution kernel to evaluate spatial locality and burst access patterns in sliding-window operations.
  \item \texttt{multihead\_attention.c}: A toy multi-head attention workload that exercises dot-product computes and softmax-induced memory reuse typical of transformer models.
  \item \texttt{trace\_example.c}: A minimal read/write trace generator used to validate basic request sequencing and correct data return in the simulator.
  \item \texttt{vector\_similarity.c}: A cosine-similarity search across a small vector database to test irregular access patterns and reduction operations.
\end{itemize}

\section{Results}

\subsection{Overview}
\label{sec:results-overview}

\begin{table}[htbp]
    \centering
    \small
    \resizebox{\linewidth}{!}{%
    \begin{tabular}{@{}lcccc@{}}
        \toprule
        \textbf{Benchmark Name} & \textbf{Read Diff Avg} & \textbf{Read StdDev} & \textbf{Write Diff Avg} & \textbf{Write StdDev} \\
        \midrule
        \texttt{conv2d.c} & 102 & 59 & 171 & 154 \\
        \texttt{multihead-attention.c} & 114 & 67 & 110 & 38 \\
        \texttt{trace-example.c} & 117 & 70 & 111 & 38 \\
        \texttt{exp-vector-similarity.c} & 110 & 66 & 109 & 38 \\
        \bottomrule
    \end{tabular}
    }
    \caption{Average cycle differences in read/write requests between DRAMSim3 (ideal software simulator) and MemorySim. Values computed as $\mathbf{MemSimCycles} - \mathbf{DRAMSimCycles}$. These results are obtained by setting the $\mathbf{queueSize}$ parameter to $128$. }
    \label{tab:general_results_overview}
\end{table}

As one may reasonably expect, the RTL-level model consumes more cycles per read or write operation. On average for traces simulations runs set to $100,000$ cycles, the read-operation consumes $\frac{102+114+117+110}{4} \approx 111$ cycles, while write operations take $\frac{171+110+111+109}{4} \approx 125$ cycles. This data, listed in \autoref{tab:general_results_overview}, alongside the standard deviations, suggest large room for improvement in the precision of the accelerator. An important characteristic to keep in mind here is the parameter $\mathbf{queueSize}$, which controls the depth of all queues within the controller system. As we will characterize in time, toggling the queue depth significantly alters the read and write difference averages, bringing the differences down to the order of $50$ and $80$ cycles respectively, also reducing the standard deviation to $50$ and $10$ cycles respectively. To sum it up, the simulator's memory subsystem is highly sensitive to \textit{backpressure} from $\mathbf{reqQueue}$, which largely increases as $\mathbf{queueSize}$ rises.

For further analysis, we demonstrate how critical of an impact back-pressure can play  (1) the mean cycles MemorySim takes over different cycle periods, and (2) varying the queue size by powers of two and observing the cycle differences. 

We also acknowledge feature-based limitations, which we've listed below:
\begin{itemize}
    \item \textbf{OPEN PAGE vs CLOSED PAGE policies}: There appears to be some systematic bugs in DRAMSim3 software, where schedulers would issue only READ requests, even after being placed in closed page policy, which would require the model to issue pre-charges after READ/WRITE requests to close rows, along with activates to open the row. In other words, DRAMSim3 enacted a default OPEN PAGE policy, in comparison to our CLOSED PAGE policy.
    \item \textbf{Lack of caching / request re-ordering}: The existing implementation currently doesn't support caching rows, which is implemented in the ideal simulator for read / write operations to bypass scheduling. As discussed in our section on next steps, these are set to be part of a later host of features to be released.
\end{itemize}

\section{Ablations}

\subsection{Cycles over Various Epochs}
\begin{figure}[htbp]
    \centering
    \includegraphics[width=1\linewidth]{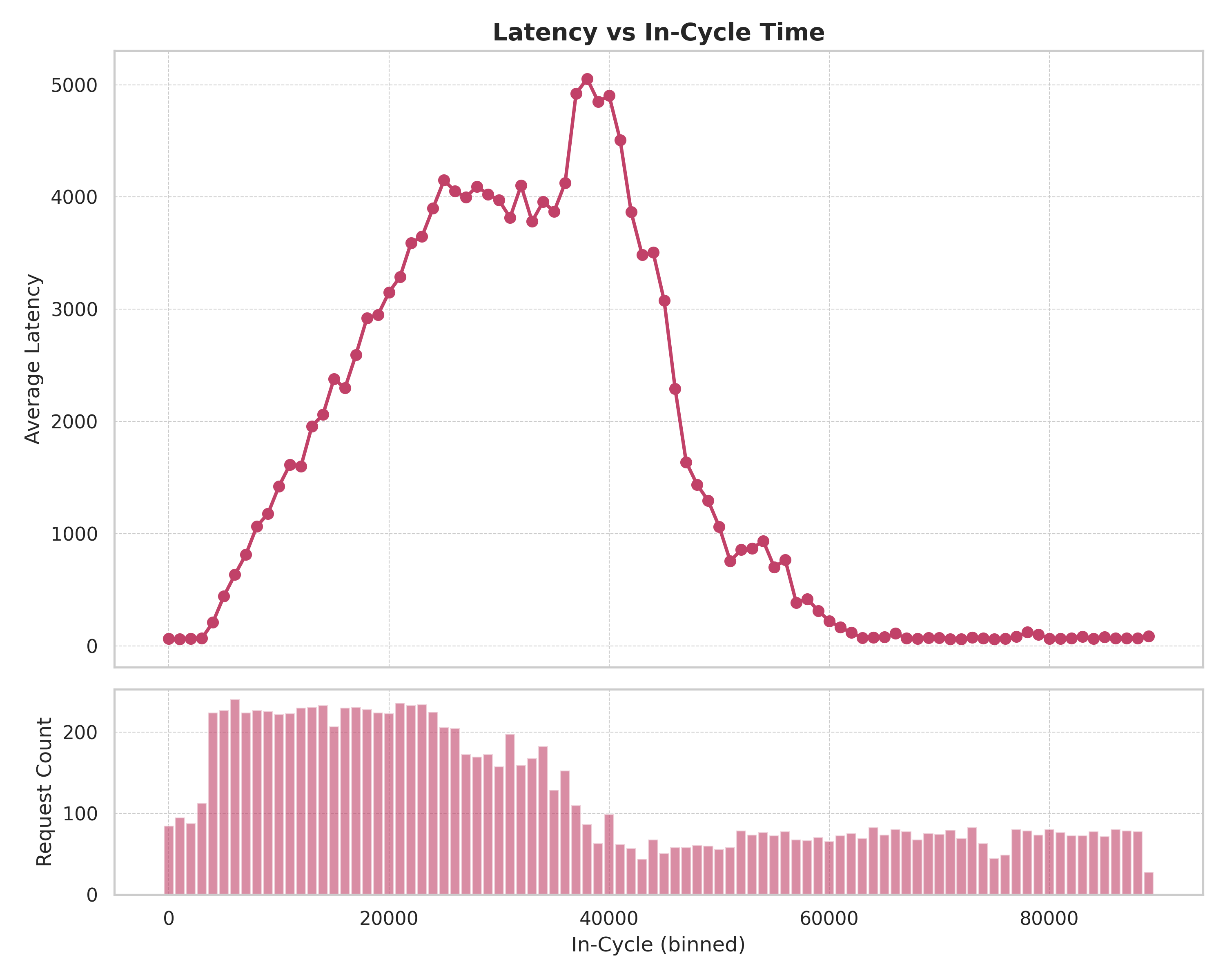}
    \caption{Simulator average latency profile, with a window of $1000$ cycles.  This was taken on the \texttt{conv2d} benchmark. }
    \label{fig:average-latency-conv2d-profile}
\end{figure}

In order to understand how the simulator is reacting to the given benchmark traces and whether peak traffic had an impact overall, we plot the average latency vs the input cycle range, i.e, the average latency of all the requests within $1000$ cycle bins. It's noticeable that request latency initially, for the first $500$ cycles, remains stable. However, as the number of cycles progress and the system faces a higher number of traffic requests, we see a rise in latency. This shows that for some initial state, the system is capable of performing, but processively is unable to tackle the load. 

\subsection{Latency vs Queue Size}
\begin{figure}[htbp]
    \centering
    \includegraphics[width=1\linewidth]{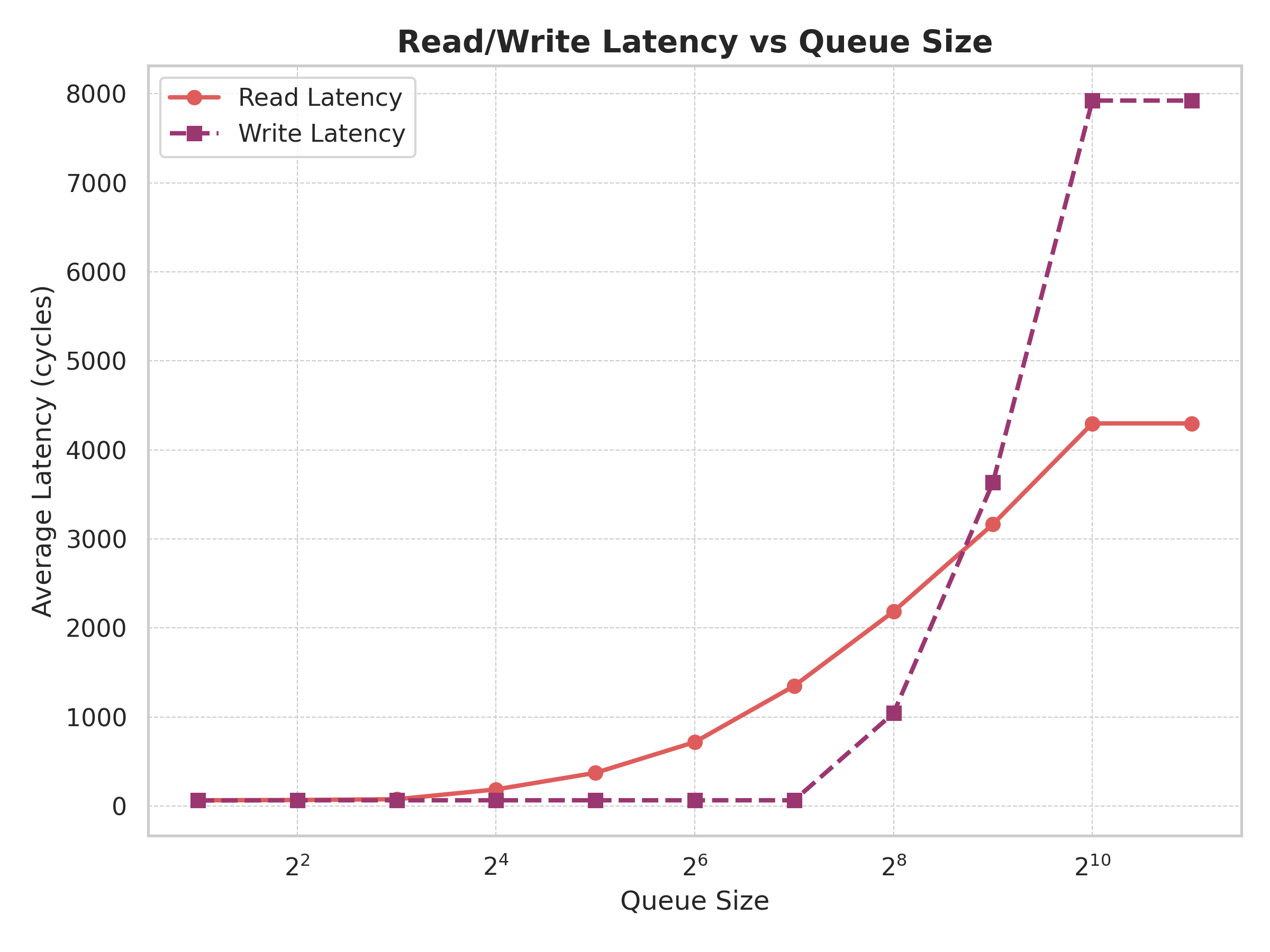}
    \caption{Read and write latency curves vs. queue size, performed on the \texttt{conv2d} benchmark. }
    \label{fig:latency-vs-queue-size}
\end{figure}

As demonstrated previously in \autoref{fig:average-latency-conv2d-profile}, traffic seems to periodically increase. That leads us to check $\mathbf{queueSize}$, the parameter that checks the size of $\mathbf{reqQueue}$ (see \autoref{fig:mem-controller-architecture} for a reminder). We vary $\mathbf{queueSize}$ between $2$ and $1024$ to examine its impact on latency. It turns out, as shown in \autoref{fig:latency-vs-queue-size}, that the latency does indeed scale with request queue size, exponentially. Decreasing the request queue size brings the latency significantly lower, while extended peak traffic activity on larger queues lead to overall larger latency.

\subsection{Simulator Cycle Breakdown}
\begin{figure}[htbp]
    \includegraphics[width=1\linewidth]{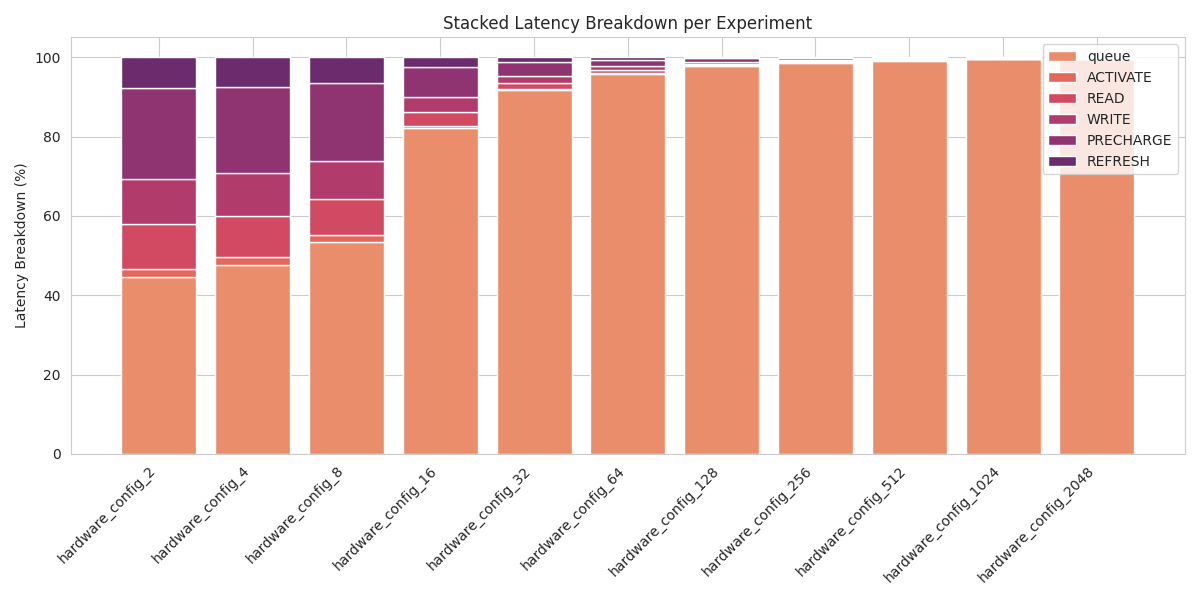}
    \caption{Breakdown of the total latencies by percentage on the \texttt{conv2d} benchmark. We vary the queue size of the controller, systematically increasing from 2 to 2048 in size. }
    \label{fig:simulator-cycle breakdown}
\end{figure}

To prove the statement that "request queue size is the single-most critical factor that determines average latency in the system", we perform a further breakdown of the average cycle latencies into their average constituents. \autoref{fig:simulator-cycle breakdown} yields an explanation into the phenomenon - Higher $\mathbf{queueSize}$ leads to degraded performance caused by increased backpressure on \textit{reqQueue}. As the size of the queues scale, the backpressure becomes all-consuming, accounting for $\approx100\%$ of the cycle delay.

\subsection{Tradeoff: Latency vs. Number of Requests}
\begin{figure}[htbp]
    \centering
    \includegraphics[width=1\linewidth]{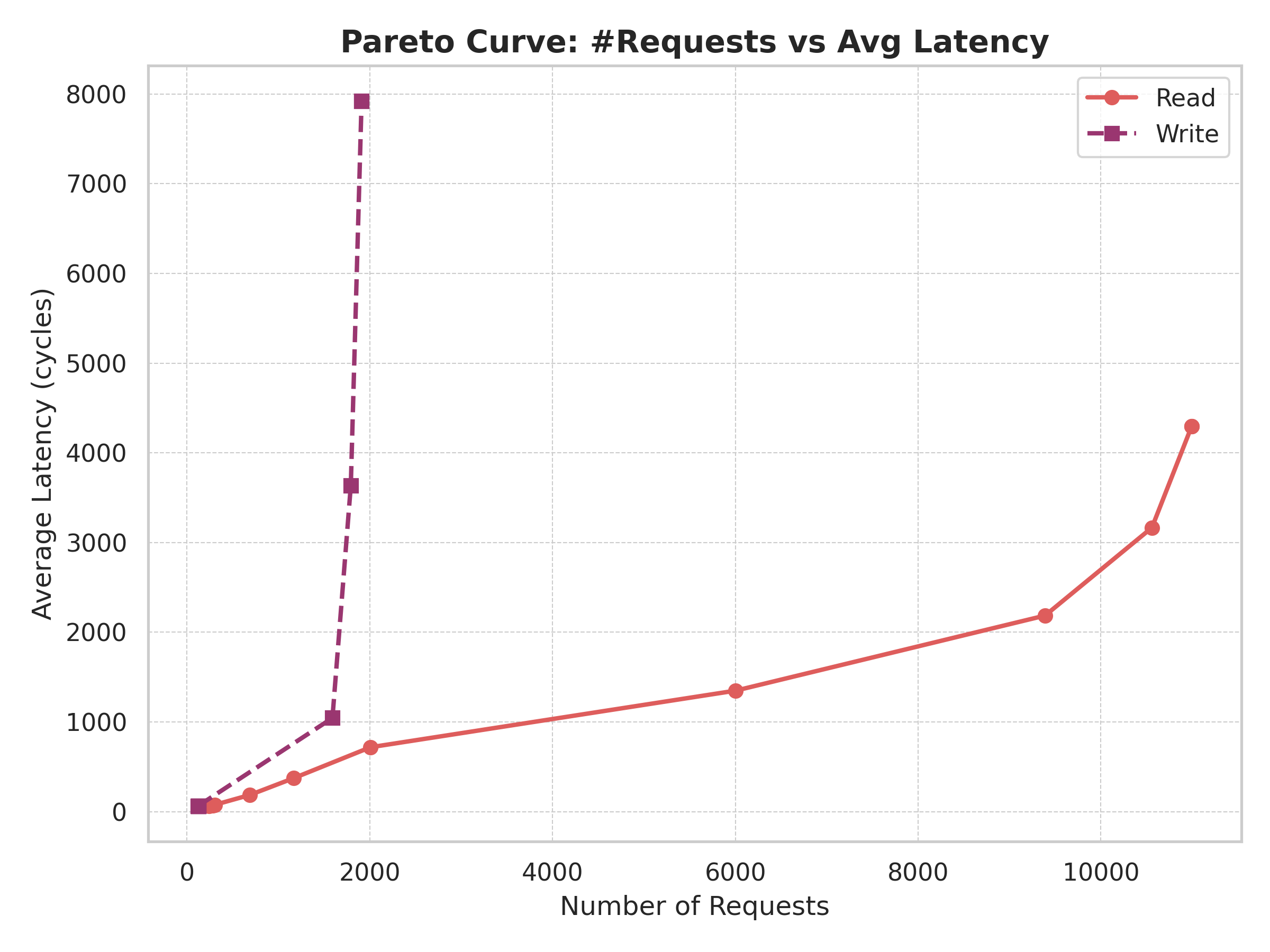}
    \caption{Pareto curve of the number of requests completed and the average latency. Ideally, we want to be at an operating point where we have high numbers of requests, and low average latency.}
    \label{fig:latency-requests-pareto}
\end{figure}

Based on the previous ablation the solution appears simple - just reduce $\mathbf{queueSize}$! Unfortunately, this fails to take into account another problem - requests that don't enter the system become \textit{blocking} for all future requests.  Consider the case where multiple bank schedulers are operating at peak capacity (full individual queues, dictated by a smaller queue size). Due to a small queue size, backpressure on the larger, request-facing queue can reach a state with requests in the queue are all backfill for a bank scheduler already operating at full capacity. As a result, all the other schedulers are starved of requests, leading to lower number of requests.

\autoref{fig:latency-requests-pareto} demonstrates the Pareto tradeoff between the number of requests served on the \texttt{conv2d} benchmark, vs. latency. Notably, the lowest average latency systems face this "starvation" problem more often, leading to lower throughput. On the other hand, higher throughput systems are more capable of unblocking themselves from this problem, at the cost at higher perceived average latency from $\mathbf{reqQueue}$ backpressure.

\section{Conclusion}

In this work, we have introduced \textit{MemorySim}, an RTL‐native DRAM simulator fully embedded within the Chisel/Chipyard ecosystem. By implementing bank‐level FSMs, a centralized \texttt{reqQueue} with multiple dequeue support, and a cycle‐accurate DRAM timing model entirely in hardware, MemorySim eliminates synthesis challenges with high level behavioural memory models.

Our quantitative evaluation against DRAMSim3 reveals a consistent cycle‐count overhead across four AI‐accelerator microbenchmarks. As shown in Table \ref{tab:general_results_overview}, MemorySim incurs an average read‐cycle penalty of $111$ cycles and a write‐cycle penalty of $125$ cycles over $100,000$ cycle traces. Notably, \texttt{conv2d.c} exhibits the highest write overhead ($171$ cycles), while \texttt{multihead\_attention.c} shows the lowest write variance (StdDev $38$) (\autoref{tab:general_results_overview}).

A detailed latency profiling (\autoref{fig:average-latency-conv2d-profile}) demonstrates that average request latency remains near $110$ cycles for the first $500$ cycles, but climbs to over $200$ cycles under sustained traffic bursts—highlighting sensitivity to transient backpressure. Varying the \texttt{queueSize} parameter between $2$ and $1024$ (\autoref{fig:latency-vs-queue-size}) yields an exponential latency increase: a queue size of 2 achieves sub-80 cycle average latency, whereas a size of $1024$ pushes latency beyond $250$ cycles.

Figure \ref{fig:simulator-cycle breakdown} breaks down the sources of latency and confirms that backpressure in the centralized \texttt{reqQueue} accounts for up to $100\%$ of additional cycles as queue depth increases. The Pareto trade‐off between throughput and latency (\autoref{fig:latency-requests-pareto}) further illustrates that minimizing \texttt{queueSize} reduces latency but can starve bank schedulers—dropping completed requests from over $10,000$ to under $6,000$ on the \texttt{conv2d} benchmark.

In summary, MemorySim delivers a high‐fidelity, modular platform for RTL‐level memory subsystem exploration. By correlating specific latency overheads with system parameters (\autoref{tab:general_results_overview}, \autoref{fig:latency-vs-queue-size}), designers can quantitatively evaluate and tune memory‐controller policies. Future work will extend these findings with dynamic backpressure controls, per‐bank read caching, and rank‐level self‐refresh optimizations to further close the performance gap to software models while retaining hardware correctness.  

\section{Code}
All reference code used can be found at this GitHub repository: \url{https://github.com/AnshKetchum/hbm-controller}.

\section{Acknowledgements}
We thank Professor Chistopher Fletcher from UC Berkeley and Ph.D student Tianrui Wei for their generous, insightful counsel.

\bibliographystyle{ACM-Reference-Format}
\bibliography{references}


\begin{thebibliography}{21}

%%% the Web address.

\ifx \showCODEN    \undefined \def \showCODEN     #1{\unskip}     \fi
\ifx \showDOI      \undefined \def \showDOI       #1{#1}\fi
\ifx \showISBNx    \undefined \def \showISBNx     #1{\unskip}     \fi
\ifx \showISBNxiii \undefined \def \showISBNxiii  #1{\unskip}     \fi
\ifx \showISSN     \undefined \def \showISSN      #1{\unskip}     \fi
\ifx \showLCCN     \undefined \def \showLCCN      #1{\unskip}     \fi
\ifx \shownote     \undefined \def \shownote      #1{#1}          \fi
\ifx \showarticletitle \undefined \def \showarticletitle #1{#1}   \fi
\ifx \showURL      \undefined \def \showURL       {\relax}        \fi
% invisible to TeX
\providecommand\bibfield[2]{#2}
\providecommand\bibinfo[2]{#2}
\providecommand\natexlab[1]{#1}
\providecommand\showeprint[2][]{arXiv:#2}

\bibitem[Ardakani et~al\mbox{.}(2017)]%
        {ardakani2017vampire}
\bibfield{author}{\bibinfo{person}{Hadi Ardakani}, \bibinfo{person}{Yuke Yang}, \bibinfo{person}{Guangyu Dong}, \bibinfo{person}{James Drake}, {and} \bibinfo{person}{Michael~J. Irwin}.} \bibinfo{year}{2017}\natexlab{}.
\newblock \showarticletitle{VAMPIRE: A DRAM Power and Performance Modeling Framework}. In \bibinfo{booktitle}{\emph{Proceedings of the 2017 IEEE International Symposium on Performance Analysis of Systems and Software}}. \bibinfo{pages}{1--12}.
\newblock


\bibitem[Arnold et~al\mbox{.}(2015)]%
        {arnold2015validating}
\bibfield{author}{\bibinfo{person}{Klaus Arnold}, \bibinfo{person}{Jin-Soo Jeong}, {and} \bibinfo{person}{David Chau}.} \bibinfo{year}{2015}\natexlab{}.
\newblock \showarticletitle{Validating DRAM Timing and Functionality in an FPGA-Based Emulation Platform}. In \bibinfo{booktitle}{\emph{Proceedings of the 2015 IEEE International Symposium on Performance Analysis of Systems and Software}}. \bibinfo{pages}{15--25}.
\newblock


\bibitem[Bachrach et~al\mbox{.}(2012)]%
        {bachrach2012chisel}
\bibfield{author}{\bibinfo{person}{Jonathan Bachrach}, \bibinfo{person}{Jonathan Vo}, \bibinfo{person}{Charles Richards}, \bibinfo{person}{Andrew Lee}, \bibinfo{person}{Andrew Waterman}, \bibinfo{person}{David Patterson}, {and} \bibinfo{person}{Krste Asanović}.} \bibinfo{year}{2012}\natexlab{}.
\newblock \showarticletitle{Chisel: Constructing Hardware in a Scala Embedded Language}. In \bibinfo{booktitle}{\emph{Proceedings of the 49th Annual Design Automation Conference}}. \bibinfo{pages}{1216--1225}.
\newblock


\bibitem[Bommasani et~al\mbox{.}(2021)]%
        {bommasani2021opportunities}
\bibfield{author}{\bibinfo{person}{Rishi Bommasani}, \bibinfo{person}{Drew~A. Hudson}, \bibinfo{person}{Ehsan Adeli}, \bibinfo{person}{Russ Altman}, \bibinfo{person}{Simran Arora}, \bibinfo{person}{Sydney von Arx}, \bibinfo{person}{Michael~S. Bernstein}, {et~al\mbox{.}}} \bibinfo{year}{2021}\natexlab{}.
\newblock \showarticletitle{On the Opportunities and Risks of Foundation Models}.
\newblock \bibinfo{journal}{\emph{arXiv preprint arXiv:2108.07258}} (\bibinfo{year}{2021}).
\newblock
\urldef\tempurl%
\url{https://arxiv.org/abs/2108.07258}
\showURL{%
\tempurl}


\bibitem[Brown et~al\mbox{.}(2020)]%
        {brown2020language}
\bibfield{author}{\bibinfo{person}{Tom~B. Brown}, \bibinfo{person}{Benjamin Mann}, \bibinfo{person}{Nick Ryder}, \bibinfo{person}{Melanie Subbiah}, \bibinfo{person}{Jared Kaplan}, \bibinfo{person}{Prafulla Dhariwal}, \bibinfo{person}{Arvind Neelakantan}, \bibinfo{person}{Pranav Shyam}, \bibinfo{person}{Girish Sastry}, \bibinfo{person}{Amanda Askell}, {et~al\mbox{.}}} \bibinfo{year}{2020}\natexlab{}.
\newblock \showarticletitle{Language Models are Few-Shot Learners}.
\newblock \bibinfo{journal}{\emph{Advances in Neural Information Processing Systems (NeurIPS)}}  \bibinfo{volume}{33} (\bibinfo{year}{2020}), \bibinfo{pages}{1877--1901}.
\newblock
\urldef\tempurl%
\url{https://arxiv.org/abs/2005.14165}
\showURL{%
\tempurl}


\bibitem[Harrison et~al\mbox{.}(2020)]%
        {harrison2020midas}
\bibfield{author}{\bibinfo{person}{Al Harrison}, \bibinfo{person}{Maria Vegdahl}, \bibinfo{person}{Nayan Parihar}, {et~al\mbox{.}}} \bibinfo{year}{2020}\natexlab{}.
\newblock \showarticletitle{MIDAS: FPGA-Accelerated Full-System Simulation at Near-Native Speed}. In \bibinfo{booktitle}{\emph{Proceedings of the 2020 IEEE International Symposium on Performance Analysis of Systems and Software}}. \bibinfo{pages}{1--12}.
\newblock


\bibitem[Kai et~al\mbox{.}(2018)]%
        {kai2018dramsys}
\bibfield{author}{\bibinfo{person}{William Kai}, \bibinfo{person}{Wim Heirman}, {and} \bibinfo{person}{Dirk Stroobandt}.} \bibinfo{year}{2018}\natexlab{}.
\newblock \showarticletitle{DRAMSys: Flexible and Extensible Open-Source Memory Subsystem Modeling in SystemC}. In \bibinfo{booktitle}{\emph{Proceedings of the 2018 IEEE International Symposium on Performance Analysis of Systems and Software}}. \bibinfo{pages}{1--11}.
\newblock


\bibitem[Kaplan et~al\mbox{.}(2020)]%
        {kaplan2020scaling}
\bibfield{author}{\bibinfo{person}{Jared Kaplan}, \bibinfo{person}{Sam McCandlish}, \bibinfo{person}{Tom Henighan}, \bibinfo{person}{Tom Brown}, \bibinfo{person}{Benjamin Chess}, \bibinfo{person}{Rewon Child}, \bibinfo{person}{Scott Gray}, \bibinfo{person}{Alec Radford}, \bibinfo{person}{Jeffrey Wu}, {and} \bibinfo{person}{Dario Amodei}.} \bibinfo{year}{2020}\natexlab{}.
\newblock \showarticletitle{Scaling Laws for Neural Language Models}.
\newblock \bibinfo{journal}{\emph{arXiv preprint arXiv:2001.08361}} (\bibinfo{year}{2020}).
\newblock
\urldef\tempurl%
\url{https://arxiv.org/abs/2001.08361}
\showURL{%
\tempurl}


\bibitem[Kim et~al\mbox{.}(2015)]%
        {kim2015ramulator}
\bibfield{author}{\bibinfo{person}{Yunji Kim}, \bibinfo{person}{Jayadev Das}, \bibinfo{person}{Onur Mutlu}, {and} \bibinfo{person}{Yale~N. Patt}.} \bibinfo{year}{2015}\natexlab{}.
\newblock \showarticletitle{Ramulator: A Fast and Extensible DRAM Simulator}. In \bibinfo{booktitle}{\emph{IEEE Computer Architecture Letters}}, Vol.~\bibinfo{volume}{15}. \bibinfo{pages}{45--49}.
\newblock


\bibitem[Kundu et~al\mbox{.}(2024)]%
        {kundu2024performance}
\bibfield{author}{\bibinfo{person}{Joyjit Kundu}, \bibinfo{person}{Wenzhe Guo}, \bibinfo{person}{Ali BanaGozar}, \bibinfo{person}{Udari De~Alwis}, \bibinfo{person}{Sourav Sengupta}, \bibinfo{person}{Puneet Gupta}, {and} \bibinfo{person}{Arindam Mallik}.} \bibinfo{year}{2024}\natexlab{}.
\newblock \showarticletitle{Performance Modeling and Workload Analysis of Distributed Large Language Model Training and Inference}.
\newblock \bibinfo{journal}{\emph{arXiv preprint arXiv:2407.14645}} (\bibinfo{year}{2024}).
\newblock
\urldef\tempurl%
\url{https://arxiv.org/abs/2407.14645}
\showURL{%
\tempurl}


\bibitem[Mosbeck et~al\mbox{.}(2015)]%
        {mosbeck2015usimm}
\bibfield{author}{\bibinfo{person}{Jakob Mosbeck}, \bibinfo{person}{Guillaume Helion}, \bibinfo{person}{Robert Dreslinski}, {and} \bibinfo{person}{et al.}} \bibinfo{year}{2015}\natexlab{}.
\newblock \showarticletitle{USIMM: A Stateful DRAM System Simulation Infrastructure}. In \bibinfo{booktitle}{\emph{Proceedings of the 2015 IEEE International Symposium on Performance Analysis of Systems and Software}}. \bibinfo{pages}{1--10}.
\newblock


\bibitem[Ning et~al\mbox{.}(2024)]%
        {ning2024survey}
\bibfield{author}{\bibinfo{person}{Xuefei Ning}, \bibinfo{person}{Yu Wang}, \bibinfo{person}{Guohao Dai}, \bibinfo{person}{Jie Xu}, \bibinfo{person}{X.-P. Zhang}, \bibinfo{person}{Yuhan Dong}, \bibinfo{person}{Zulin Yuan}, \bibinfo{person}{Shujie Yan}, {and} \bibinfo{person}{Xiaoyang Li}.} \bibinfo{year}{2024}\natexlab{}.
\newblock \showarticletitle{A Survey on Efficient Inference for Large Language Models}.
\newblock \bibinfo{journal}{\emph{arXiv preprint arXiv:2404.14294}} (\bibinfo{year}{2024}).
\newblock
\urldef\tempurl%
\url{https://arxiv.org/abs/2404.14294}
\showURL{%
\tempurl}


\bibitem[Rajbhandari et~al\mbox{.}(2019)]%
        {rajbhandari2020zero}
\bibfield{author}{\bibinfo{person}{Samyam Rajbhandari}, \bibinfo{person}{Jeff Rasley}, \bibinfo{person}{Olatunji Ruwase}, {and} \bibinfo{person}{Yuxiong He}.} \bibinfo{year}{2019}\natexlab{}.
\newblock \showarticletitle{ZeRO: Memory Optimizations Toward Training Trillion Parameter Models}.
\newblock \bibinfo{journal}{\emph{arXiv preprint arXiv:1910.02054}} (\bibinfo{year}{2019}).
\newblock
\urldef\tempurl%
\url{https://arxiv.org/abs/1910.02054}
\showURL{%
\tempurl}


\bibitem[Recasens et~al\mbox{.}(2025)]%
        {recasens2025mind}
\bibfield{author}{\bibinfo{person}{Pol~G. Recasens}, \bibinfo{person}{Ferran Agullo}, \bibinfo{person}{Yue Zhu}, \bibinfo{person}{Chen Wang}, \bibinfo{person}{Eun~Kyung Lee}, \bibinfo{person}{Olivier Tardieu}, \bibinfo{person}{Jordi Torres}, {and} \bibinfo{person}{Josep~Ll. Berral}.} \bibinfo{year}{2025}\natexlab{}.
\newblock \showarticletitle{Mind the Memory Gap: Unveiling GPU Bottlenecks in Large-Batch LLM Inference}.
\newblock \bibinfo{journal}{\emph{arXiv preprint arXiv:2503.08311}} (\bibinfo{year}{2025}).
\newblock
\urldef\tempurl%
\url{https://arxiv.org/abs/2503.08311}
\showURL{%
\tempurl}


\bibitem[Schellekens et~al\mbox{.}(2014)]%
        {schellekens2014dram}
\bibfield{author}{\bibinfo{person}{Wim Schellekens}, \bibinfo{person}{Sam Swinnen}, {and} \bibinfo{person}{Dirk Stroobandt}.} \bibinfo{year}{2014}\natexlab{}.
\newblock \showarticletitle{{DRAMPower}: Open-Source DRAM Power Energy Estimation Tool}. In \bibinfo{booktitle}{\emph{Proceedings of the 2014 IEEE International Symposium on Performance Analysis of Systems and Software}}. \bibinfo{pages}{1--12}.
\newblock


\bibitem[Shevgoor et~al\mbox{.}(2016)]%
        {shevgoor2016understanding}
\bibfield{author}{\bibinfo{person}{Santhosh Shevgoor}, \bibinfo{person}{Onur Mutlu}, {and} \bibinfo{person}{Alok Choudhary}.} \bibinfo{year}{2016}\natexlab{}.
\newblock \showarticletitle{Understanding the Impact of DRAM Latency on Memory System Performance}. In \bibinfo{booktitle}{\emph{Proceedings of the 2016 ACM International Conference on Measurement and Modeling of Computer Systems}}. \bibinfo{pages}{390--401}.
\newblock


\bibitem[Vaswani et~al\mbox{.}(2017)]%
        {vaswani2017attention}
\bibfield{author}{\bibinfo{person}{Ashish Vaswani}, \bibinfo{person}{Noam Shazeer}, \bibinfo{person}{Niki Parmar}, \bibinfo{person}{Jakob Uszkoreit}, \bibinfo{person}{Llion Jones}, \bibinfo{person}{Aidan~N. Gomez}, \bibinfo{person}{{\L}ukasz Kaiser}, {and} \bibinfo{person}{Illia Polosukhin}.} \bibinfo{year}{2017}\natexlab{}.
\newblock \showarticletitle{Attention Is All You Need}. In \bibinfo{booktitle}{\emph{Advances in Neural Information Processing Systems (NeurIPS)}}. \bibinfo{pages}{5998--6008}.
\newblock
\urldef\tempurl%
\url{https://arxiv.org/abs/1706.03762}
\showURL{%
\tempurl}


\bibitem[Waterman et~al\mbox{.}(2020)]%
        {waterman2020chipyard}
\bibfield{author}{\bibinfo{person}{Andrew Waterman}, \bibinfo{person}{Krste Asanović}, \bibinfo{person}{David Patterson}, {and} \bibinfo{person}{Jonathan Bachrach}.} \bibinfo{year}{2020}\natexlab{}.
\newblock \showarticletitle{Chipyard: Integrated Design, Simulation, and Implementation Framework for Custom SoCs}.
\newblock \bibinfo{journal}{\emph{arXiv preprint arXiv:2001.00046}} (\bibinfo{year}{2020}).
\newblock


\bibitem[Weber et~al\mbox{.}(2010)]%
        {weber2010dramsim2}
\bibfield{author}{\bibinfo{person}{Henry Weber}, \bibinfo{person}{Bavani Weber}, {and} \bibinfo{person}{Per Stenstrom}.} \bibinfo{year}{2010}\natexlab{}.
\newblock \showarticletitle{{DRAMSim2}: A Cycle Accurate Memory System Simulator}. In \bibinfo{booktitle}{\emph{Proceedings of the 2010 43rd Annual IEEE/ACM International Symposium on Microarchitecture}}. \bibinfo{pages}{2--13}.
\newblock


\bibitem[Yu et~al\mbox{.}(2021)]%
        {yu2021dramsim3}
\bibfield{author}{\bibinfo{person}{Chen Yu}, \bibinfo{person}{Aparna Chandrasekar}, \bibinfo{person}{Vivek Seshadri}, {et~al\mbox{.}}} \bibinfo{year}{2021}\natexlab{}.
\newblock \showarticletitle{DRAMSim3: A Memory System Simulator}.
\newblock \bibinfo{journal}{\emph{arXiv preprint arXiv:2102.05607}} (\bibinfo{year}{2021}).
\newblock


\bibitem[Zhang et~al\mbox{.}(2020)]%
        {zhang2020firesim}
\bibfield{author}{\bibinfo{person}{Andrew Zhang}, \bibinfo{person}{Jacob Izraelevitz}, \bibinfo{person}{Onur Guney}, {et~al\mbox{.}}} \bibinfo{year}{2020}\natexlab{}.
\newblock \showarticletitle{FireSim: FPGA-Accelerated Cycle-Exact Scale-Out System Simulation in the Public Cloud}. In \bibinfo{booktitle}{\emph{Proceedings of the 2020 USENIX Annual Technical Conference}}. \bibinfo{pages}{1--15}.
\newblock


\end{thebibliography}
\end{document}